%% file: draft.tex
\documentclass[10pt]{revtex4}

\usepackage{amsmath}
\usepackage{amssymb}
\usepackage{lineno}

\input{myDefinitions}

\usepackage{graphicx}
\usepackage[parfill]{parskip}
\begin{document}

\title{Periodic fluctuations in deep water formation due to sea ice}

\author{Raj Saha}
\affiliation{University of Minnesota, Department of Mathematics\\Mathematics and Climate Research Network}
\email{rsaha@umn.edu}

\begin{abstract}
\input{abstract}
\end{abstract}

\maketitle


\input{sec01}
\input{sec02}

\input{sec03}

\input{sec04}
\input{sec05}
\input{sec06}

\begin{acknowledgements}
I would like to thank Chris Jones and John Bane for advising my research project. In addition I would like to thank NSF and the Mathematics and Climate Research Network for support.
\end{acknowledgements}


%
%

\include{bibliography}
\bibliography{bibliography}
\bibliographystyle{plain}


\include{bibliography}  
\end{document}

%% file: myDefinitions.tex
\newcommand\T{\rule{0pt}{2.6ex}}

\newcommand{\ben}{\begin{enumerate}}
\newcommand{\een}{\end{enumerate}}

\newcommand{\be}{\begin{equation}}
\newcommand{\ee}{\end{equation}}
\newcommand{\ba}{\begin{eqnarray}}
\newcommand{\ea}{\end{eqnarray}}

\newcommand{\bmat}{\left(\begin{array}}
\newcommand{\emat}{\end{array}\right)}

  \newcommand{\e}{\epsilon}

  \newcommand{\G}{\Gamma}

\renewcommand{\deg}{$^{\circ}$}

\newcommand{\nete}{$\varepsilon$}


%% file: abstract.tex
During the last ice age several quasi-periodic abrupt warming events took place. Known as Dansgaard-Oeschger (DO) events their effects were felt globally, although the North Atlantic experienced the largest temperature anomalies. Paleoclimate data shows that the fluctuations often occurred right after massive glacial meltwater releases in the North Atlantic and in bursts of three or four with progressively decreasing strengths. In this study a simple dynamical model of an overturning circulation and sea ice is developed with the goal of understanding the fundamental mechanisms that could have caused the DO events. Interaction between sea ice and the overturning circulation in the model produces self-sustained oscillations. Analysis and numerical experiments reveal that the  insulating effect of sea ice causes the ocean to periodically vent out accumulated heat in the deep ocean into the atmosphere. Subjecting the model to idealized freshwater forcing mimicking Heinrich events causes modulation of the natural periodicity and produces burst patterns very similar to what is observed in temperature proxy data. Numerical experiments with the model also suggests that the characteristic period of 1,500 years is due to the geometry, or the effective heat capacity, of the ocean that comes under sea ice cover.

%% file: sec01.tex
\section{Introduction}

\begin{figure}[htbp]
\begin{center}
	\includegraphics[width=5in]{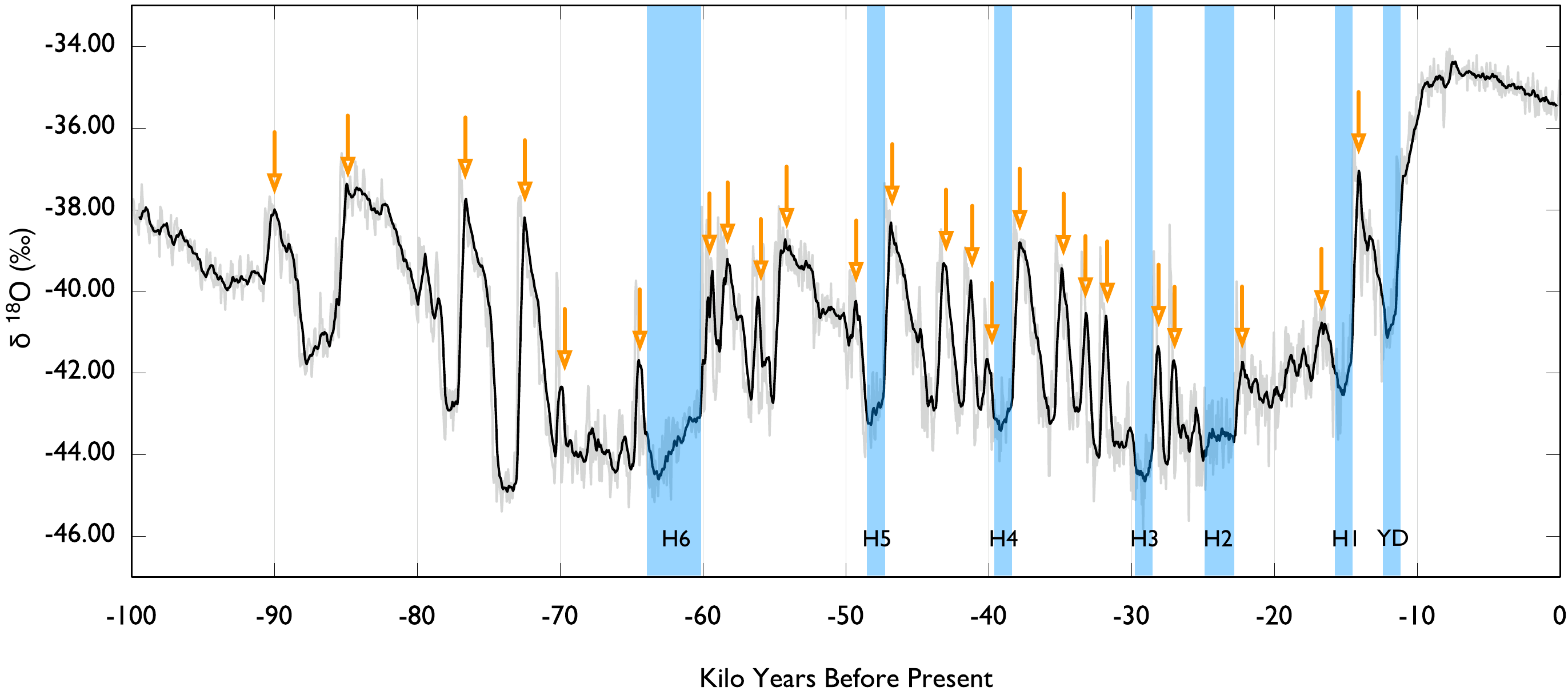}
\caption{Oxygen-18 isotopic concentrations in Greenland ice cores (NGRIP) showing the variation of near surface air temperatures over the last 100,000 years. The blue vertical stripes indicate times when Heinrich events took place.}
\label{fig:DO}
\end{center}
\end{figure}

Climate proxy data from ice \cite{dansgaard1993} and sediment \cite{wang2001} cores reveal that the climate of the north Atlantic underwent large quasi periodic fluctuations during the last glacial period (figure \ref{fig:DO}). These are the so called Dansgaard-Oeschger (DO) events and they occurred in intervals of about 1,500 years. Fluctuations on a similar timescale but with lesser intensity are also seen during the Holocene warm period \cite{bond1997}. This paper proposes a mechanism whereby sea ice interacts with oceanic overturning circulation to generate millennial scale oscillations in climate. A simple dynamical model is used to show how the self-sustained oscillations could be modulated by large freshwater perturbations to produce temporal patterns resembling DO events.

DO events were characterized by large and abrupt warming of the north Atlantic followed by slower cooling. The proxy records indicate that the strength of the warming anomaly decreased in magnitude farther out from Greenland and had the opposite sign in the southern hemisphere. This bipolar see-saw behavior suggests a global reorganization of heat flow \cite{stocker2003} via changes in oceanic circulation. Conservation of the total amount heat in the climate system means that a regional warming must be balanced by cooling elsewhere. In the current climate state a convective region exists in the Norwegian sea which connects to the larger oceanic `conveyor' through bottom flow. Several studies \cite{rahmstorf_2002,hofmann_rahmstorf_2009} state that DO events were the direct result of fluctuations in deep water formation in the north Atlantic.

Some DO events were preceded by massive floods that released glacial meltwater into the north Atlantic (Heinrich events). Model studies have demonstrated that large freshwater additions could temporarily amplify the convection to produce abrupt warming. However only some DO events can be directly attributed to Heinrich events as there are far fewer of the latter. The mismatch in the timescales of DO and Heinrich events suggest the existence of a separate mechanism.

A number of mechanisms have been proposed to explain the source of the millennial scale oscillations. Some suggest astronomical forcing such as solar output \cite{braun2005}, or lunar tidal resonance \cite{keeling2000}. Some others suggest internal mechanisms due to periodic variability in convective strength. However by themselves these mechanisms do not explain many of the observed characteristics of DO events. For example proxies measuring solar output do not show any correlation with the temperature records (ref). The lunar tidal resonance hypothesis, which connects tidal resonances to the destabilization of ice sheets, do not explain the variability in the pacing of DO events. The internal variability mechanisms of `deep-decoupling' \cite{lenderink1994,winton_1993} are highly sensitive to  freshwater perturbations and are unlikely to persist over long durations and under different climate states. A stable oscillation mechanism is required to explain DO events which persisted all through the glacial period and possibly through the Holocene as well.

In this paper a sea ice-convection climate oscillator is proposed as a solution. Some studies have noted that sea ice could be an ingredient in convective instabilities. A sea ice oscillator model was proposed by \citep{saltzman2002} that worked by sea ice regulating the ocean's uptake of $CO_2$. While atmospheric $CO_2$ variations are unlikely to be the cause for DO events, for reasons discussed earlier, sea ice could act as a regulator for the deep ocean's heat storage \cite{rial_yang_2007}. The insulating property of sea ice, its seasonal recurrence and the geophysical constraints on its spatial extent might be at the heart of a persistent millennial scale oscillation. The proceeding sections describe a simple dynamical 'box' model of sea ice interacting with an overturning circulation. Box models are useful tools in studies of this nature as their simplicity allows for analysis and elucidation of the fundamental ingredients of a climatic phenomenon. 

%% file: sec02.tex
\section{A simple dynamical model}
\label{sec:model}
The model is constructed by coupling an existing model of thermodynamic sea ice \cite{GT2001b} to an ocean box model \cite{verdiere2006}. The ocean model consists of boxes arranged into three meridional zones and two vertical layers. Advective transport between the boxes is driven by horizontal pressure gradients in the top mixed layer where temperature and salinity determine the density of sea water in each box. Sea ice forms or melts according to the temperature of the polar box and it's extent is limited to the bounds of the top polar box. The main way sea ice interacts with the ocean is by cutting off heat exchange with the atmosphere. These two models are among the simplest available examples of the respective processes.

Some modifications are introduced for each component. Sea ice is assumed to be a perfect insulator, unlike \cite{verdiere2006} where it had non zero heat permeability. In the case of the ocean model the masses of individual boxes are computed and not kept fixed as in \cite{verdiere2006}. This seemingly unnecessary modification is needed to avoid an unphysical situation to arise out of the convective mechanism, namely  two parcels of equally dense waters mixing uniformly and end up with a slightly higher density. This one-time step density anomaly leads to a large jump in pressure gradients. The resulting anomaly in advective flux destabilizes the system and produces relaxation oscillations. Such a scenario however is not physical and is an artifact of the discretized nature of the model. Upon computing the box masses runtime, the 'free-oscillations' reported in \cite{verdiere2006} disappear and only steady states are observed. Another modification was made to the ocean model by removing diffusive mixing as its inclusion was found to not be necessary in generating oscillations.

\subsection{Governing equations}
The system is defined by a set of ODEs accounting for the rate of change of heat and salt content of each box. Thermal and salinity (freshwater) forcing is applied to the surface boxes. The equations are:
\begin{eqnarray}\label{eq:goveq}
	m_i C_{\mathrm{p}} \frac{dT_i}{dt}&=& Q_i A_i + \rho_o C_{\mathrm{p}} \psi_{i,j} T_j + \mathrm{Co}_T(T_i, S_i, T_{i+3}, S_{i+3})  + \phi_T(T_3) \\
	m_i \frac{dS_i}{dt}&=& F_i + \psi_{i,j} S_j + \mathrm{Co}_S(T_i, S_i, T_{i+3}, S_{i+3}) + \phi_S(T_3)  \\
	\frac{df}{dt}&=&k_1(T_0-T_3)+k_2 f
\end{eqnarray}
where the left-hand side of the equations are the rates of change of heat and salinity for a box of index $i$. The same numbering scheme is used for the boxes as in \cite{verdiere2006}. Terms on the right-hand side of the equations are individual components of heat or salinity flux from a box to its neighboring boxes, the air above it, or from the latent heat and brine rejection from sea ice formation. The following sections explain each term and the associated processes. 

\subsection{Geometry}
The ocean boxes occupy three meridional zones and two vertical layers. The southernmost zone spans between 45$^{\circ}$ and 60$^{\circ}$N, the intermediate zone between 60$^{\circ}$ and 75$^{\circ}$N, and the polar zone between 75$^{\circ}$ and 90$^{\circ}$N. The vertical layers are 1000 m, representing the top mixed layer, and 3500 m in depth. Areas of the vertical faces between boxes are computed according to a spherical shell geometry of the ocean. 

Sea ice is restricted to the top polar box only. It is prescribed a thickness of 2 m. However, as will be described in forthcoming sections, the volume of sea ice does not contribute to the model's oscillation.

\subsection{Salinity (freshwater) forcing}
On a given surface box, evaporation and precipitation is modeled as a prescribed rate of change of salinity. 
\begin{equation}
	F_i=\rho_o S_o \epsilon A_1 q_i
\end{equation}
where $\epsilon$ is the net tropical evaporation amount in depth per unit time, $A_i$ is the surface area of the tropical box, and $q_i$ are the fractional E-P proportions prescribed as
\begin{eqnarray}
	q_1&=&+1\\
	q_2&=&-0.7\\
	q_3&=&-0.3
\end{eqnarray}

The top equatorial box experiences net evaporation, so $F_1>0$, while the other surface boxes experience net precipitation so that $F_2, F_3 <0$. The E-P proportions are such that the total amount of salt in the system is conserved. 

\subsection{Thermal forcing}
The surface boxes are exposed to air temperatures on a prescribed gradient, with the pole being colder than the equator. For sub-polar surface boxes, the atmospheric heat exchange is given by
\begin{equation}
	Q_i = \lambda (\Gamma_i-T_i)A_i
\end{equation}
where $\lambda$ is a prescribed rate of heat transfer per unit area, $\Gamma_i$ is the air temperature above box i, and $A_i$ is the area of contact with air. Sea ice on the polar box will vary the surface area of contact, so that
\begin{equation}
	Q_3=\lambda (\Gamma_i-T_3)A_3 (1-f)
\end{equation}
where $f$ is the fraction of the polar area covered by sea ice. 

\subsection{Advection}
Advective flows in the model are driven by horizontal pressure gradients and are given by
\begin{equation}
	\psi_{i,i+1}=C A_{i,i+1}(P_i-P_{i+1})
\end{equation}
where the parameter C is the inverse of a friction coefficient whose value of chosen to give appropriate flow rates for the North Atlantic, $A_{i,i+1}$ is the area of contact between adjoining boxes, and $P_i, P_{i+1}$ are the respective pressures in the boxes. Pressure is formulated in a way such that the net pressure averaged across a vertical column is zero. 
\begin{eqnarray}
	P_i&=&-\frac{1}{2} \frac{h_{i+3}}{h_{i}+h_{i+3}} (h_{i} \rho_{i}+h_{i+3} \rho_{i+3})g \\
	P_{i+3}&=&-\frac{h_{i}}{h_{i+3}} P_i
\end{eqnarray}
where $h_{i}$ and $h_{i+3}$ are the depths, and $\rho_{i}$ and $\rho_{i+3}$ are the densities of a pair of vertically stacked top and bottom boxes. Density is computed from a linear equation of state given as
\begin{equation}
	\rho_{i}=\rho_o(1-\alpha (T_i-T_r)+\beta (S_i-S_r))
\end{equation}
where $\alpha$ and $\beta$ are thermal expansion and haline contraction coefficients respectively, and $T_r, S_r$ are reference temperature and salinity values. 

Volume conservation constraints require the upwelling term between boxes 2 and 5 to be
\begin{equation}
	\psi_w=\psi_{2,3}-\psi_{1,2}
\end{equation}

The flow constraints also require equalization of top and bottom flows, leading to
\begin{eqnarray}
	\psi_{1,2}=\psi_{5,4}\\
	\psi_{2,3}=\psi_{6,5}
\end{eqnarray}		
A positive value of $\psi_{1,2}$ or $\psi_{2,3}$ implies poleward flow on the surface. Thus for a given box, the net advective heat exchange is given as the sum of flows going through it. For the purpose of illustration if $\psi_{1,2}>0$, then the advective heat budget for box 1 is given by
\begin{equation}	
	-\rho_o C_{\mathrm{p}}\psi_{1,2}(T_1-T_2)+\rho_o C_{\mathrm{p}}\psi_{1,2}(T_4-T_1)
\end{equation}
The flow terms are generalized using the upstream difference scheme following \cite{thual_mcwilliams_1992}, and a 6x6 advection matrix is constructed. Elements of this matrix are given in the appendix. The advection of salinity follows a similar approach.

\subsection{Convection}
In the event of a surface box becoming more dense than the one below it, the heat and salt contents of the two boxes are mixed evenly in one time step. The equalized temperature and salinity values are given by
\begin{eqnarray}
	\bar{S}&=& \frac{m_i S_i+m_{i+3} S_{i+3}}{m_i+m_{i+3}}\\
	\bar{T}&=& \frac{m_i T_i+m_{i+3} T_{i+3}}{m_i+m_{i+3}}
\end{eqnarray}		
Thus convection as modeled is either on or off depending on polar stratification

\subsection{Sea ice}
Sea ice forms/melts when the temperature of the surface polar box goes below/above freezing. The governing growth rate by volume is given by
\begin{equation}
	\frac{dV_\mathrm{sea-ice}}{dt}=\frac{\rho_o C_{\mathrm{p}}V_3}{\rho_{\mathrm{sea-ice}}L_f \tau}(T_0 - T_3) + f p_3
\end{equation}
where $V_3$ is the volume of the polar box, $\tau$ is a prescribed timescale for sea ice growth - about 3 months, $T_0$ is the freezing temperature, and $p_3$ is the precipitation rate by volume. The second term turns precipitation on existing sea ice into ice.  The remaining physical constants are described in Table \ref{table:parameters}.

As sea ice forms it is spread with uniform thickness on the polar box. The rate of change in fractional sea ice area coverage is then given by
\begin{equation}
	\frac{df}{dt}=\frac{1}{d_{\mathrm{sea-ice}} A_3}\frac{dV_\mathrm{sea-ice}}{dt}
\end{equation}
where $d_{\mathrm{sea-ice}}=2$m, and $A_3$ is the total area of the polar box. This becomes another governing equation of the system in addition to eqs (\ref{eq:goveq}). In the case where the entire polar box is covered, any additional ice volume goes towards increasing the thickness of ice. 

The formation and melting of sea ice has enthalpy of heat and brine rejection associated with it. The enthalpy of heat is given by
\begin{equation}
	\phi_T(T_3)=\frac{\rho_o C_{\mathrm{p}}V_3}{\tau}(\Gamma_3-T_3)
\end{equation}

The salinity adjustment from the excess salt left behind from sea ice formation (brine rejection) is given by
\begin{equation}
	\phi_S(T_3)=\frac{\rho_o}{\rho_\mathrm{sea-ice}L_f}\phi_T(T_3)
\end{equation}

\begin{table}\footnotesize	
\begin{center}
\begin{tabular}{ r | l | l  }
\hline \hline
$g$ \T& Acceleration due to gravity & 9.8 m s$^{-2}$\\
$\rho_{o}$ \T& Reference density of sea water & 1028 kg m$^{-3}$\\
$\rho_{\mathrm{sea-ice}}$ \T& Density of ice & 920 kg m$^{-3}$\\
$T_{r}$ \T& Reference temperature & 283 K\\
$S_{r}$ \T& Reference salinity & 0.035\\
$\alpha$ \T& Thermal expansion coefficient & 1.7$\times 10^{-4}$ K$^{-1}$\\
$\beta$ \T& Haline contraction coefficient & 0.76 \\
$\eta$ \T& Thermal gradient & -0.74 K/\deg lat \\
\nete \T& Net evaporation from surface tropical box & 12 mm/day \\
$q_{i}$ \T& Salinity forcing ratios for surface boxes & 1, -0.7, -0.2, -0.1 \\
$c$ \T& Temperature forcing for box 1, thermal intercept & 35 \deg C\\
$T_{\mathrm{ice}}$ \T& Temperature at which ice forms & 273 K\\
$L_{e}$ \T& Latent heat due to evaporation &  2.2$\times10^{6}$ J kg$^{-1}$\\
$L_{f}$ \T& Latent heat fusion of ice & 3.34$\times10^{5}$ J kg$^{-1}$\\
$\lambda$ \T& Heat exchange coefficient between atmosphere and ocean & 30 W m$^{-2}$ K$^{-1}$\\
$C$ \T& Advective transport coefficient (inverse of friction coefficient) & $10^{-5}$ m$^{2}$ s kg$^{-1}$\\
$d_{\mathrm{sea-ice}}$ \T& Thickness of sea ice & 2 m\\
$\tau$ \T& Growth/decay timescale of sea ice& 3 months \\
$m_{i}$ \T& Mass of box $i$ & kg\\
$A_{i,j}$ \T& Cross sectional area between box $i$ and $j$& m$^{2}$\\
$T_{i}$ \T& Temperature of box $i$ & K\\
$S_{i}$ \T& Salinity of box $i$ & dimensionless\\
$\G_{i}$ \T& Surface thermal (atmospheric temperature) forcing for box $i$ & K\\
$Q_{i}$ \T& Heat exchange between atmosphere and surface box & J s$^{-1}$\\
$\psi_{i,j}$ \T& Advective volume transport between boxes $i$ and $j$ & m$^{3}$s$^{-1}$\\
$\mathit{Co}(T_{i})$ \T& Temperature adjustment due to convection & K\\
$\mathit{Co}(S_{i})$ \T& Salinity adjustment due to convection & dimensionless\\
$F_{i}$ \T& Salinity forcing on box $i$& kg m$^{-2}$ s$^{-1}$\\
$\phi_{S}$ \T& Heat exchange between sea ice and surface polar box (box 4) & J s$^{-1}$\\
$\phi_{S}$ \T& Salinity forcing due to brine rejection from sea ice formation and melting & kg s$^{-1}$ \\
\hline \hline
\end{tabular}
\label{table:parameters}
\caption{Description of model parameters with their default values and computed variables}
\end{center}	
\end{table}

%% file: sec03.tex
\section{Model dynamics}

\begin{figure}[htbp]
\begin{center}
	\includegraphics[width=5in]{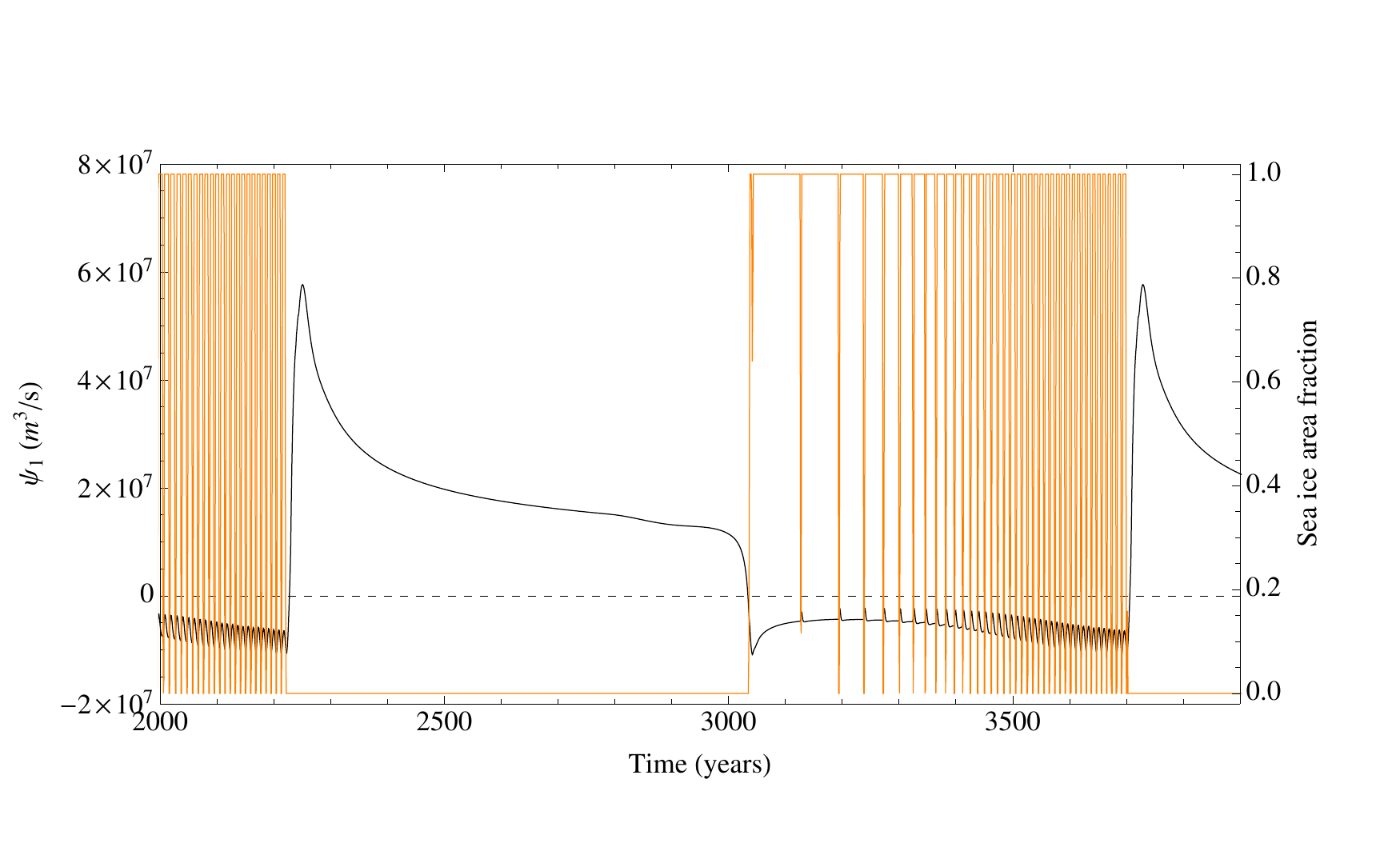}
\caption{Advective flux $\psi_1$ varying in time. }
\label{fig:oscillations}
\end{center}
\end{figure}

The model exhibits three distinct types of equilibrium flows. Two of these are steady states with opposite meridional flows. The third is an oscillation of sea ice extent, advective fluxes and convective mixing. In oscillating states the circulation periodically switches direction, alternating between small amplitude multi-decadal scale oscillations and a large amplitude multi-centennial scale fluctuation (figure (\ref{fig:oscillations})). This feature fits the description of mixed mode oscillations, the implications of which are discussed in subsection \ref{subsec:bifurcations}. 

Poleward flows in the top layer are identified as thermal (TH) flows, while flows in the opposite direction are called haline (HA). Oscillating states are denoted by OS. 

Sea ice is a necessary ingredient for oscillations. The uncoupled ocean model (with the modification described in section \ref{sec:model}) does not exhibit oscillations. 

Oscillations are characterized by an abrupt increase in poleward advective flux following the onset of polar convection. The intensified poleward advective flux and ventilation by convective mixing results in abrupt warming of the polar mixed layer ocean. The high flux state is transient and thereafter the circulation relaxes back to small amplitude oscillations with sea ice. Eventually and inevitably polar convection occurs and the process repeats in a self-sustained manner.

\begin{figure}[htbp]
\begin{center}
	\includegraphics[width=3.5in]{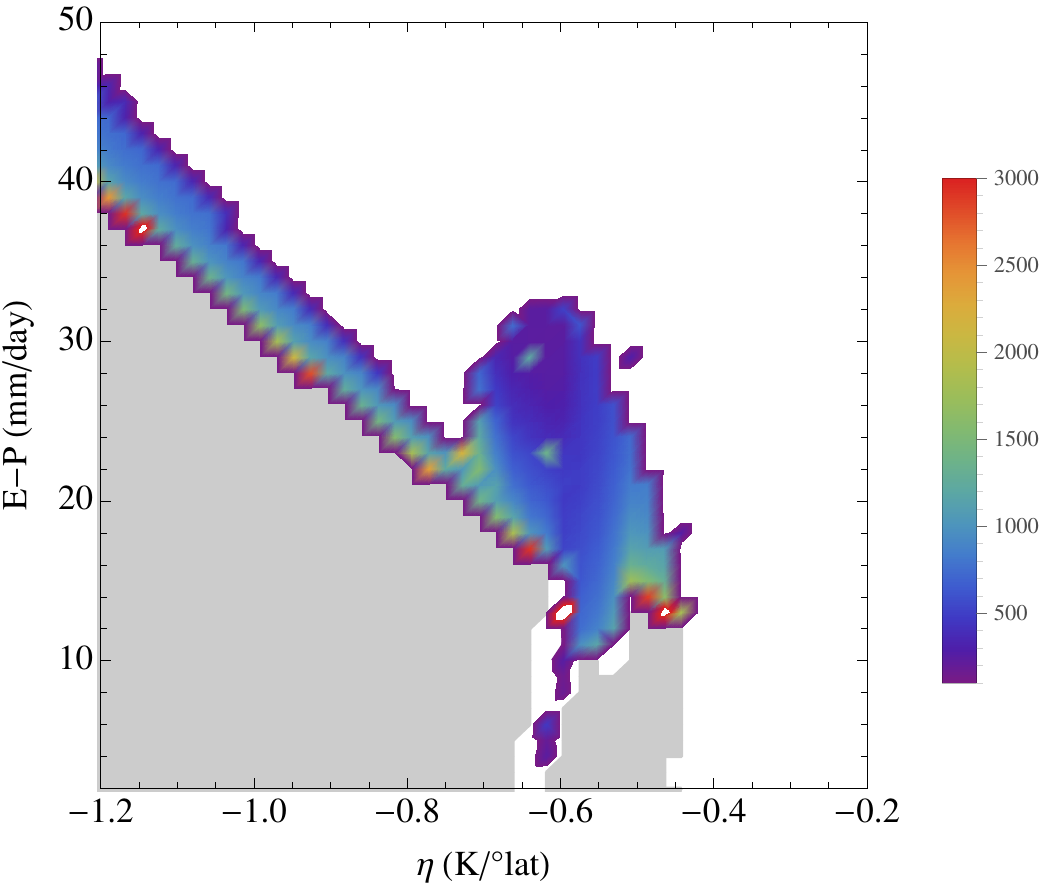}
\caption{Steady state advection ( $\psi_1$) values in the thermal and salinity forcing space, for the ocean-sea ice model. The blue line separates TH (bottom region) and HA (top region) states. OS states are enclosed within a red line, whereas polar oscillations are enclosed by a black line. Both modes of oscillations lie along approximately fixed values of $\eta/\e$.}
\label{fig:forcingspace_SI}
\end{center}
\end{figure}

The model's response is examined for a large range of thermal and freshwater forcing values. Independent simulations are carried out with 100x100 pairs of forcing values within the two-dimensional forcing space of specified size. Time series from each simulation is analyzed to detect equilibrium fluxes, or oscillation frequency and amplitude as the case may be. Figure (\ref{fig:forcingspace_SI}) identifies the TH, OS and HA states in the forcing space. 

TH states occur when the thermal forcing is stronger relative to the salinity forcing, and the reverse is true for HA states. OS states occur along the boundary of TH and HA states where the two forcings nearly balance each other. The relaxed phase of oscillations are small amplitude weakly haline flows. 

\begin{figure}[htbp]
\begin{center}
	\includegraphics[width=3.5in]{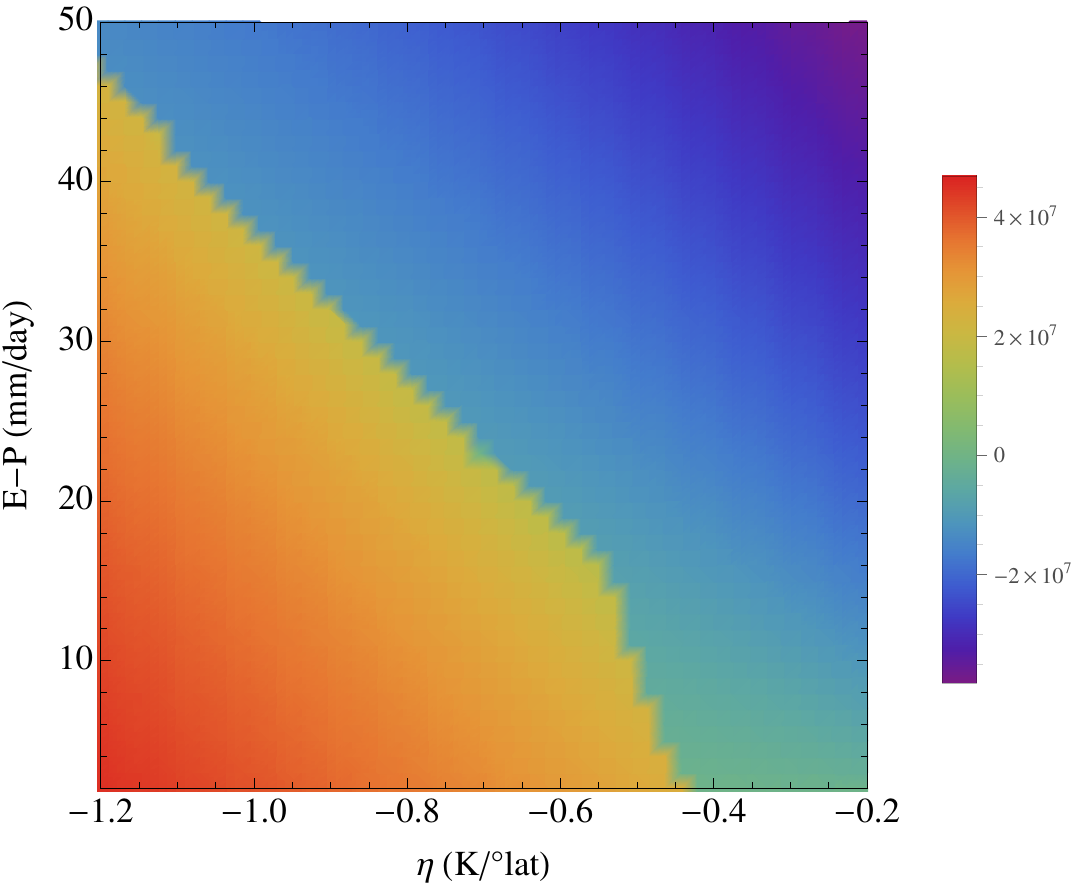}
\caption{Steady state advection ( $\psi_1$) values in the thermal and salinity forcing space, for the ocean model without sea ice. The blue line separates TH and HA states.}
\label{fig:forcingspace_noSI}
\end{center}
\end{figure}

The same tests are carried out with the decoupled ocean-only model (figure (\ref{fig:forcingspace_noSI})). No oscillations are detected and only steady states are observed.

\subsection{Bifurcations}
\label{subsec:bifurcations}
\begin{figure}[htbp]
\begin{center}
	\includegraphics[width=3.5in]{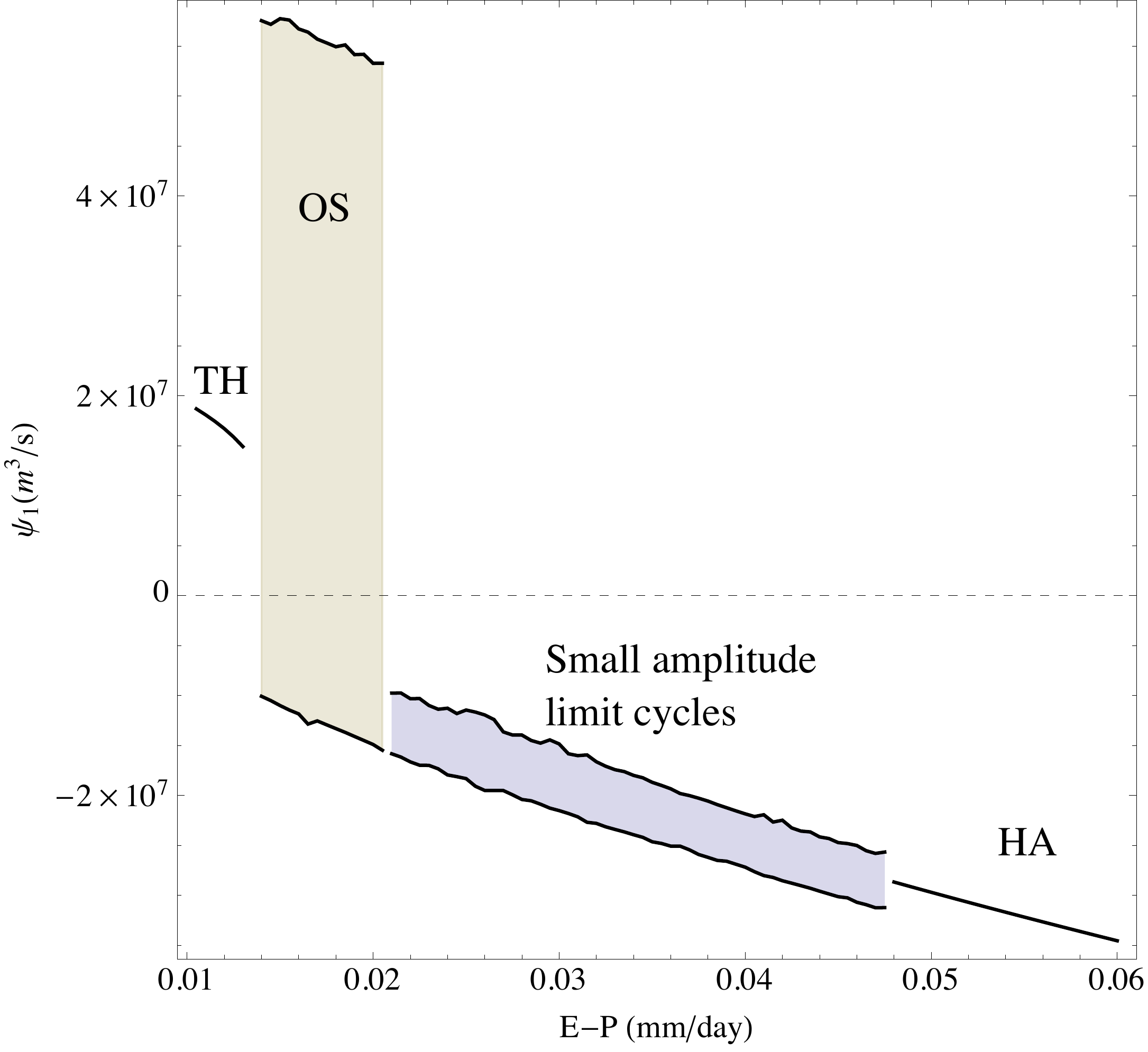}
\caption{Bifurcation diagram for slowly varying salinity forcing parameter $\e$.}
\label{fig:bifurcations}
\end{center}
\end{figure}

In order to explore the bifurcation structure of the model the thermal forcing gradient is fixed while the salinity gradient is slowly increased. The model is allowed to run for a sufficient length of time to let it come to equilibrium. If oscillations occur then the maximum and minim advective fluxes are noted. Figure (\ref{fig:bifurcations}) shows the transition from TH to OS to HA states in the direction of increasing salinity forcing.

Dynamically the TH states are weak unstable spirals. As the salinity gradient is increased the degree of instability increases until eventually the system's trajectory touches an unstable manifold of saddle (in the case of no sea ice) or a small amplitude limit cycle (with sea ice). As the minimum number of prognostic dynamical variables of the system must be three, since the oscillations are of mixed mode, the limit cycles are helical trajectories in three or higher dimensional phase space. The helical limit cycle eventually collides with the convective manifold which takes the system to a weakly unstable TH state. Under sufficiently high values of salinity forcing the limit cycle never touches the convective manifold and the small amplitude limit cycles persist. Above a threshold salinity forcing the limit cycle undergoes a supercritical Hopf bifurcation to settle on a stable spiral.

\subsection{Physical mechanism behind oscillations}
Sea ice affects the heat budget of the polar water column and causes it to periodically ventilate (lose) heat to the atmosphere. When sea ice cover is present it blocks heat loss from the mixed layer and causes its temperature to rise. This eventually causes the sea ice to melt 

Rapid convective mixing of the polar boxes drive the system away from stability and at the same time allow the density anomalies to equalize. The circulation then relaxes back to a small amplitude limit cycle which eventually lead to convection, and the process repeats.

%% file: sec04.tex
\section{Period dependence on parameters}
Parameters that control the oscillation periods are the ratio of thermal to salinity forcing, the growth/decay timescale for sea ice and the overturning advective flux rate constant. For a selected parameter the model's behavior is analyzed over a sufficiently large range of values and steady and oscillatory states are identified.

\subsection{Ratio of thermal to salinity forcing magnitudes}
\begin{figure}[htbp]
\begin{center}
	\includegraphics[width=3.5in]{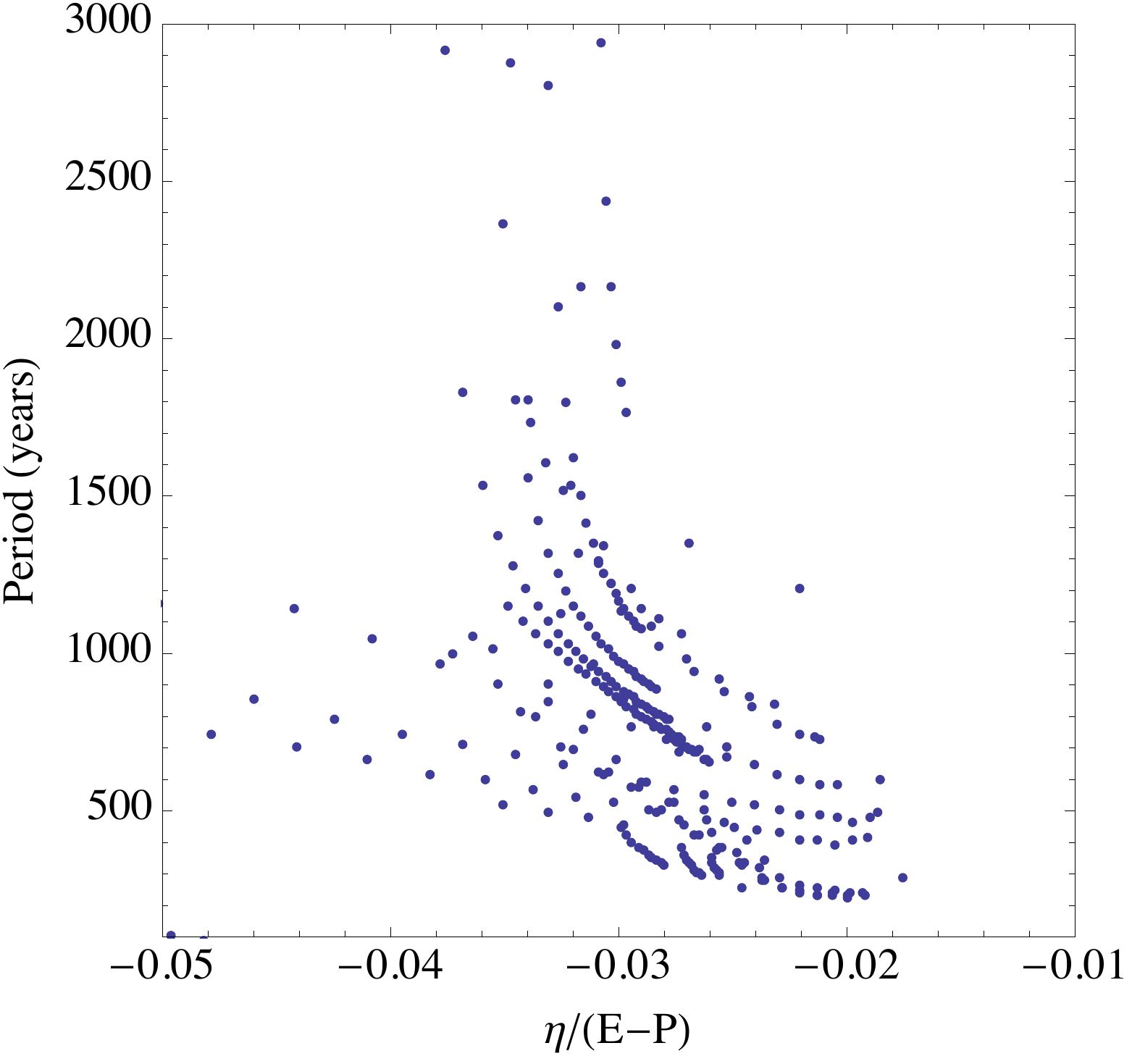}
\caption{Distribution of periods over the ratio of forcing strength parameters. Larger negative values of the ratio imply greater relative strength of thermal forcing.}
\label{fig:periods-etaemp}
\end{center}
\end{figure}

In each oscillation cycle there are two phases, first where the instabilities grow leading up to convection and a strong TH flow, followed by relaxation to a weak HA state. In general longer oscillation periods are seen to correspond to a stronger thermal forcing relative to salinity (figure \ref{fig:periods-etaemp}) because the system is forced against relaxation. Specific forcing values also determine the periodicity as they affect both the relaxation and instability building timescales.

\subsection{Timescale for sea ice growth/decay}
The growth and melt timescale for sea ice depends on a large number of physical conditions in the real world, such as the spatial and seasonal variations in near surface temperatures, ocean-atmosphere heat fluxes and strength of currents. In the model this timescale is prescribed to be in the 2-3 year range, which is largely an educated guess of how fast sea ice could grow over the Nordic seas if the background climate were similar to the last glacial period. During LGM winter sea ice extended south of Iceland, whereas today its southern extent is up to Svalbard.  The area of sea between Svalbard and Iceland is considerably larger than the present seasonal variations in sea ice extent. Considering the annual seasonal cycle of growth and melt the default timescale for sea ice is reasonable. 

\begin{figure}[htbp]
\begin{center}
	\includegraphics[width=3.5in]{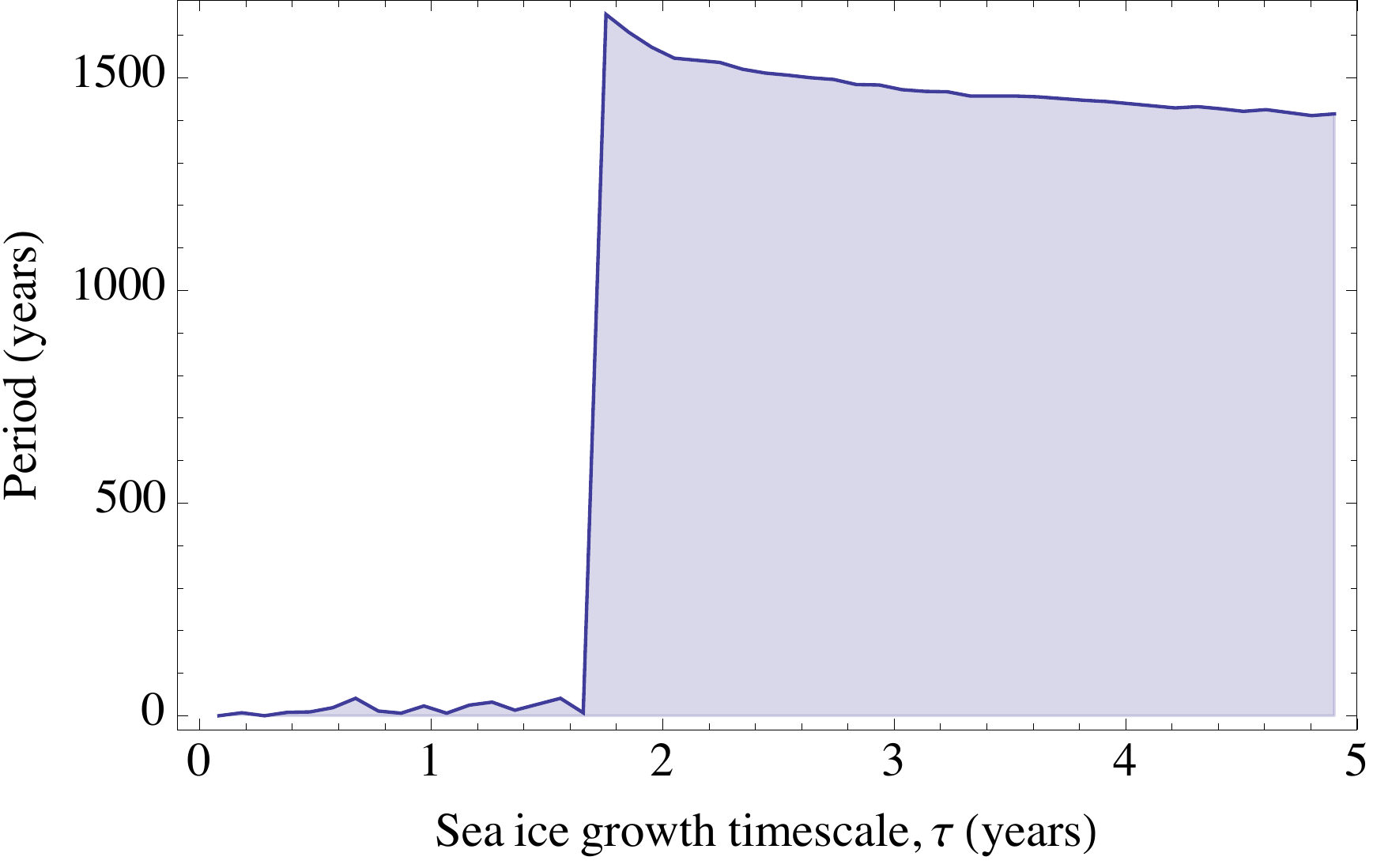}
\caption{Distribution of periods over the varying timescales of sea ice growth.}
\label{fig:periods-tauSI}
\end{center}
\end{figure}

With all other parameters fixed at their default values (Table \ref{table:parameters}), the model is run for a range of sea ice timescales, from a few months to several years. An abrupt transition occurs around a value of 1.5 years, above which oscillation periods approach 1,400 years (figure \ref{fig:periods-tauSI}). The threshold occurs where the insulating effect of sea ice in trapping heat in the ocean is matched by the dissipative heat loss by conduction and advection. As there is a maximum limit to the extent of sea ice, which in the real world is constrained by continental land masses and surface ocean currents, the timescale for heat ventilation does not vary further for longer sea ice growth timescales.

\subsection{Advective flux rate constant}
\begin{figure}[htbp]
\begin{center}
	\includegraphics[width=3.5in]{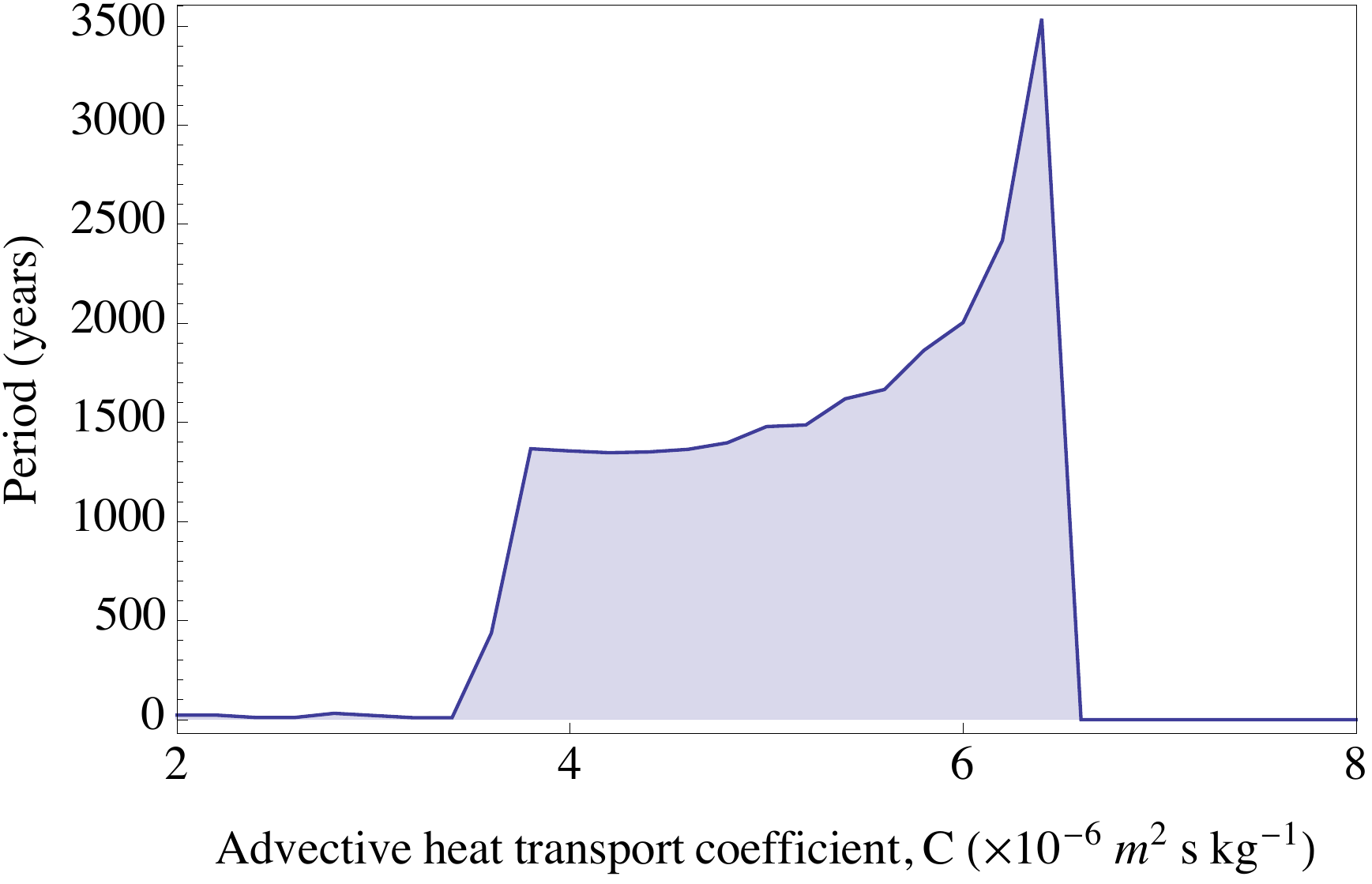}
\caption{Distribution of periods over the strength of advective flow rate.}
\label{fig:periods-C}
\end{center}
\end{figure}

The parameter $C$ prescribes the meridional flow rate per unit pressure gradient in the upper mixed layer. Its default value is based on current flow rates in the North Atlantic \cite{verdiere2006}. However this rate was likely very different between glacial and interglacial times \cite{okazaki2010}. In the context of a relaxation oscillator, especially one where the flow is conserved, the rate of flow would influence timescales of both instability and relaxation. As a result the model should exhibit oscillations only within a range of flow rates. For small flow rates the heat build up in the ocean depths would fail to bring about convective instability, whereas high flow rates would equalize density anomalies efficiently and prevent oscillations. This understanding is motivated by the model simulation results over a range of flow rate values (figure \ref{fig:periods-C}). 

%% file: sec05.tex
\section{Modulation by freshwater forcing}

The addition of freshwater to the ocean surface due to catastrophic collapse of ice sheets has been speculated to be the trigger for DO events. Simulations with intermediate complexity models do show that freshwater additions in the North Atlantic can abruptly destabilize and enhance the meridional circulation [refs]. The higher complexity models however do not seem to capture an internal climatic oscillator and consequently the model responses do not exhibit periodic behavior in the absence of periodic freshwater forcing. When periodic freshwater pulses, mimicking Heinrich events, are applied, the models respond with state switching on the same timescale as the forcing. The climate data shows that multiple DO events typically occurred in between Heinrich events with no known freshwater triggers on the smaller timescale.

The effect of freshwater on the circulation state alone is examined by adding incremental amounts of freshwater to box 1, with no sea ice present on the polar box. The freshwater forcing is varied slowly, allowing the model to come to equilibrium in each step. A hysteresis is observed when the variation in freshwater additions is reversed, showing a bi-stable region and a stable HA mode (figure). In the absence of sea ice there is no mechanism for the circulation to undergo relaxation oscillations and consequently those aren't observed.

A freshwater signal is constructed to simulate the model's response to Heinrich events occurring in intervals of 7,500 years. A gaussian function is used to represent the catastrophic freshwater pulse, and a linearly increasing freshwater addition rate is made to represent the increased basal melting leading up to a Heinrich event. Sea ice coupling is present in these simulations as the goal is to observe the effect of periodic forcing on the oscillator.

For appropriate forcing values the model produces a modulated response. Each Heinrich event triggers a series of progressively weaker abrupt warming (figure), similar to what is observed in the climate record. The gradual addition of freshwater to the surface in between Heinrich events decreases the meridional pressure gradient and results in weaker TH states. The freshwater pulse abruptly ceases poleward circulation which initiates sea ice growth, which in turn causes a resumption of convection and poleward circulation.

Variation of high latitude summer insolation during the last glacial period was least in the window of 35 to 50 ky before present. This is the same time window where DO events occurred with the regularity simulated with the model, possibly due to minimal influence of insolation forcing in that period. Speculating further it could be stated that DO events were most pronounced during the glacial period, in comparison to the Holocene, because the background circulation was in a weaker state. A combination of solar and surface freshwater forcing could explain the climatic fluctuations during the last 100 ky, which is out of the scope of this paper and largely a diagnosis of the system.

\begin{figure}[htbp]
\begin{center}
	\includegraphics[width=4in]{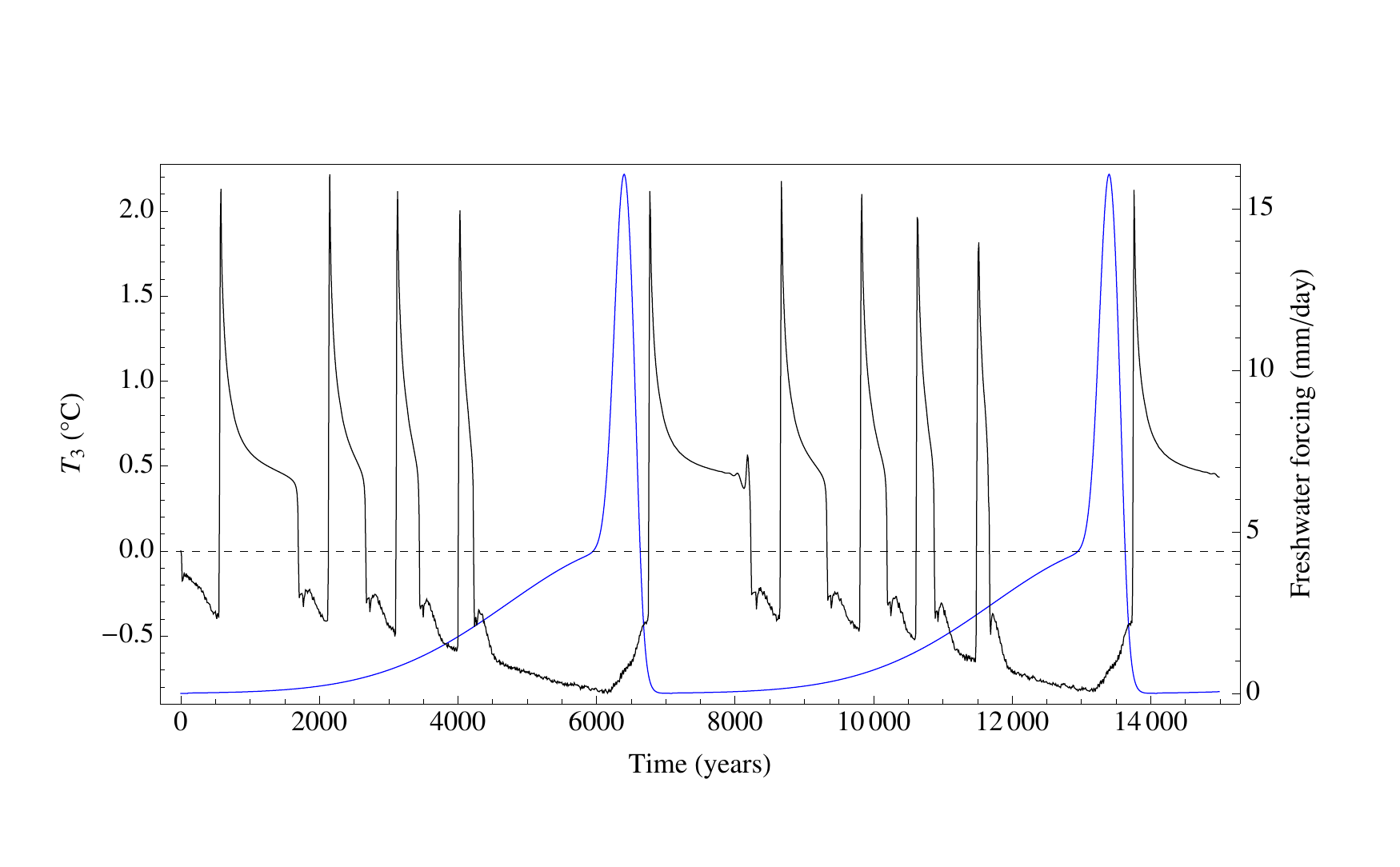}
\caption{Simulation of Heinrich events that produce temperature fluctuation patterns similar to observations in ice cores.}
\label{fig:periods-etaemp}
\end{center}
\end{figure}

%% file: sec06.tex
\section{Discussion and conclusions}

A simple dynamical model is used to show that sea ice and convection in the north Atlantic could interact to produce millennial scale quasi-periodic fluctuations. In the idealized model the fluctuations are stable periodic orbits under a large range of climatic forcing. The timescale of model oscillations and the temporal pattern of variability is very similar to that of DO events and Bond Cycles. Time varying freshwater anomalies due to ice sheet discharge and variations in obliquity modulates the intrinsic periodic response of the sea ice-convection oscillator. This oscillator could be at the heart of the millennial scale climatic variability known to have occurred during the last 100,000 years.

The oscillation is due a periodic build up of convective instability in the polar ocean. Sea ice cover when present insulates the ocean from losing heat to the atmosphere. However over repeated cycles of sea ice advance and retreat the upper mixed oceanic layer undergoes a net loss of heat due to the large heat ventilation during retreat phases. The gradual increase in mixed layer density leads to convective instability which is released by abrupt increase in vertical mixing in the polar water column. The high mixing rate removes the density anomalies and allows the circulation to relax back to its preferred state of small amplitude oscillations in sea ice advance and retreat. From here on the net positive heat loss from the mixed layer helps build up the convective anomaly and repeat the process in a self-sustained manner. 

Dynamically the oscillations are composed of two components operating on a decadal and millennial timescale respectively. The small amplitude decadal scale oscillations are weakly unstable spiral orbits approaching a limit cycle. If the trajectory intersects with the convective manifold then a non-smooth transition takes place and the system traverses to an unstable high mixing state. The mixed mode character of the sea ice oscillator means that there are at least three principal dynamical variables that are necessary ingredients. These are likely the horizontal and vertical density gradients in the north Atlantic ocean, and the extent of sea ice. Could it be that the well observed Atlantic Multi-decadal oscillations are the small amplitude component of a mixed mode oscillation of the sea ice-convection system?

The period of oscillations in the model are dependent on the climatic forcing as well as physical attributes of the system. Sea ice extent is geophysically and climatologically limited within the corridor of Greenland-Icelandic-Norwegian seas and this could set the intrinsic timescale of 1,500 years. Freshwater pulses mimicking Heinrich events modulate the periodic response into repeating groups of successively weaker and shorter fluctuations. This response is similar to the Greenland temperature proxy signal in the window of 30,000 to 50,000 years before present. In this time window the summer insolation at high latitudes had very low variability as compared to the rest of the glacial period and thus surface freshwater forcing may have been the dominating influence. 

A box model in spite of its simplicity is an effective tool in identifying the fundamental mechanisms and ingredients behind the climatic phenomenon addressed in this paper. The persistent fluctuations occurring through both glacial and interglacial periods hint at a robust oscillation mechanism for which the sea ice-convection oscillator is a suitable candidate. New mathematical techniques to study non-smooth dynamical systems such as this one is needed for further exploration of the dynamics of this climate oscillator.

%% file: draft.bbl
\begin{thebibliography}{10}

\bibitem{bond1997}
G.~Bond, W.~Showers, M.~Cheseby, R.~Lotti, P.~Almasi, P.~deMenocal, P.~Priore,
  H.~Cullen, I.~Hajdas, and G.~Bonani.
\newblock A pervasive millennial-scale cycle in north atlantic holocene and
  glacial climates.
\newblock {\em Science}, 278:1257, 1266 1997.

\bibitem{braun2005}
H.~Braun, M.~Christl, S.~Rahmstorf, A.~Ganopolski, A.~Mangini, C.~Kubatzki,
  K.~Roth, and B.~Kromer.
\newblock Possible solar origin of the 1,470-year glacial climate cycle
  demonstrated in a coupled model.
\newblock {\em Nature}, 438:208--211, 2005.

\bibitem{dansgaard1993}
W.~Dansgaard.
\newblock Evidence for general instability of past climate from a 250-kyr
  ice-core record.
\newblock {\em Nature}, 364:218--220, 1993.

\bibitem{verdiere2006}
C.~de~Verdi\`{e}re, M.~B. Jelloul, and F.~Sevellec.
\newblock Bifurcation structure of thermohaline millennial oscillations.
\newblock {\em Journal of Climate}, 19:5777--5795, 2006.

\bibitem{GT2001b}
H.~Gildor and E.~Tziperman.
\newblock A sea ice climate switch mechanism for the 100-kyr glacial cycles.
\newblock {\em Journal of Geophysical Research}, 106(C5):9117, 9133 2001.

\bibitem{hofmann_rahmstorf_2009}
M.~Hofmann and S.~Rahmstorf.
\newblock On the stability of the atlantic meridional overturning circulation.
\newblock {\em PNAS}, 10(1073), 2009.

\bibitem{keeling2000}
C.~D. Keeling and T.~P. Whorf.
\newblock The 1,800-year oceanic tidal cycle: A possible cause of rapid climate
  change.
\newblock {\em Proceedings of the Natural Academy of Sciences (PNAS)},
  97:3814--3819, April 2000.

\bibitem{lenderink1994}
G.~Lenderink and R.~J. Haarsma.
\newblock Variability and miltiple equilibria of the thermohaline circulation
  associated with deep-water formation.
\newblock {\em Journal of Physical Oceanography}, 24:1480--1493, 1994.

\bibitem{okazaki2010}
Y.~Okazaki, A.~Timmermann, L.~Menviel, N.~Harada, A.~Abe-Ouchi, M.~O.
  Chikamoto, A.~Mouchet, and H.~Asahi.
\newblock Deepwater formation in the {N}orth {P}acific during the last glacial
  termination.
\newblock {\em Science}, 329:200--204, 2010.

\bibitem{rahmstorf_2002}
S.~Rahmstorf.
\newblock Ocean circulation and climate during the past 120,000 years.
\newblock {\em Nature}, 419:207--214, 2002.

\bibitem{rial_yang_2007}
J.~A. Rial and M.~Yang.
\newblock Is abrupt climate change paced by the orbital insolation?
\newblock {\em Geophysical Monograph}, 173:167--174, 2007.

\bibitem{saltzman2002}
B.~Saltzman.
\newblock {\em Dynamical Paleoclimatology: Generalized Theory of Global Climate
  Change}.
\newblock 2002.

\bibitem{stocker2003}
T.~Stocker and S.~J. Johnsen.
\newblock A minimum thermodynamic model for the bipolar seesaw.
\newblock {\em Paleooceanography}, 18(4), 2003.

\bibitem{thual_mcwilliams_1992}
O.~Thual and J.~McWilliams.
\newblock The catastrophe structure of thermohaline convection in a
  two-dimensional fluid model and a comparison with low-order box models.
\newblock {\em Geophysical \& Astrophysical Fluid Dynamics}, 64:67--95, 1992.

\bibitem{wang2001}
Y.~J. Wang, H.~Cheng, R.~L. Edwards, Z.~S. An, J.~Y. Wu, C.-C Shen, and J.~A.
  Dorale.
\newblock A high-resolution absolute-dated late pleistocene monsoon record from
  {H}ulu cave, {C}hina.
\newblock {\em Science}, 294:2345--2348, 2001.

\bibitem{winton_1993}
M.~Winton and E.~S. Sarachik.
\newblock Thermohaline oscillations induced by strong steady stalinity forcing
  of ocean general circulation models.
\newblock {\em Journal of Physical Oceanography}, 23(7):1389--1410, 1993.

\end{thebibliography}
